\documentclass{epl}
\institute{
\inst{1} Physics Department and Center for Stochastic
Processes in Science and Engineering, 
Virginia Polytechnic Institute and State University,
Blacksburg, VA. 24061-0435, USA \\
\inst{2} Lehrstuhl Werkstoffkunde und Technologie der Metalle (WTM), 
Universit\"{a}t Erlangen-N\"{u}rnberg, 
Martensstr.~5, 91058 Erlangen, Germany.
}


\begin{document}

\title{Vacancy-mediated domain growth in a driven lattice gas. }

\author{B. Schmittmann\inst{1} \and M. Thies\inst{2}}
\maketitle

\vspace{-.5cm}

\begin{abstract}
Using Monte Carlo simulations and a mean-field theory, we study domain growth 
in a driven lattice gas. Mediated by a single vacancy, two
species of particles (``charges'') unmix, so that a disordered initial
configuration develops charge-segregated domains which grow logarithmically
slowly. An order parameter is defined and shown to satisfy dynamic scaling.
Its scaling form is computed analytically, in excellent agreement with
simulations.
\end{abstract}

\vspace{-0.8cm} 

\emph{Introduction. }Driven diffusive lattice gases, first
introduced by Katz, et.al.~\cite{KLS}, display generic non-equilibrium
features in their whole phase space~\cite{DL17}. Even when interactions are
due to excluded volume constraints alone, complex phase diagrams can be
induced by local dynamic rules~\cite{SHZ,ABC-ss,ABC-t} which violate
detailed balance~\cite{DB}. While disordered phases and universal properties
near continuous transitions are quite well understood by now, ordered phases
have proven far more complex~\cite{DL17}. In particular, studies of phase
ordering and coarsening in \emph{driven} systems have revealed surprising
behaviors, quite distinct from those~\cite{GB} exhibited by systems evolving
towards an \emph{equilibrium} final state. For example, when driven, the
two-dimensional (2D) Ising lattice gas develops non-universal domain
morphologies which grow in a highly anisotropic fashion and do not satisfy
dynamic scaling~\cite{CAL}. In 1D, ordered domains grow with time as 
$t^{1/2}$~\cite{CB}, in contrast to the equilibrium $t^{1/3}$ law~\cite{CKS}.
Analytic results are sparse and focus primarily on establishing growth 
laws~\cite{CB,ABC-t,2D-c,2D-nc}. Clearly, further studies of domain growth far
from equilibrium are needed before a more general framework can be
established.

In this letter, we report a detailed dynamic scaling analysis for
defect-mediated domain growth in a driven three-state lattice gas~\cite{SHZ}%
. Two species of particles, labelled ``positive'' and
``negative'', diffuse on a periodic lattice by hopping onto  
a single empty site (the defect). Only nearest-neighbor 
jumps are allowed, biased by an ``electric'' field $E$ aligned with a 
lattice axis. At \emph{finite} vacancy concentration, this system
exhibits a phase transition, controlled by drive and particle density, from
a disordered, ``free-flowing'' phase to an ordered, ``jammed'' phase~\cite
{SHZ,BSZ,LZ}. Ordered configurations are spatially inhomogeneous: positive
and negative particles form adjacent strips transverse to $E$, 
impeding each other. A mean-field theory predicts the 
structure and stability limits of both phases~\cite{SHZ,LZ,VZS}.
Models of this type have been invoked to describe water-in-oil
microemulsions~\cite{waterdroplets}, gel electrophoresis~\cite{gelelectro}
and traffic flow~\cite{trafficflow}.

Remarkably, even the presence of a \emph{single} vacancy suffices to order
an initially random system. This problem, in both its static and dynamic
aspects, is an example of a much-wider ranging class of interacting random
walk and defect-mediated domain growth problems. The vacancy plays the role
of a highly mobile defect~\cite{VMD}, whose motion restructures its
environment (i.e., the particle configuration), while the latter provides a
feedback through the local jump rates. In the resulting steady state,
investigated in detail in~\cite{TS}, particles form a charge-segregated
strip, with the vacancy \emph{localized} at one of the interfaces. The
density profiles obey characteristic scaling forms, controlled by drive and
system size.

Here, we focus on the \emph{time evolution} of this system in 2D,
seeking to understand how ordered domains form,
grow and finally saturate. We present Monte Carlo (MC) data, 
supported by an approximate solution of the
time-dependent mean-field equations. The ordering process exhibits three
regimes: the initial formation of the strip, an extended
logarithmic growth regime and the crossover to steady state. We
demonstrate that the system satisfies dynamic scaling in the second
regime and compute the scaling function analytically, in excellent
agreement with the data. We conclude with some suggestions for further
study. More details can be found in~\cite{long}.

\emph{The microscopic model. }Our model is defined on a 2D square lattice of 
$L_{x}\times L_{y}$ sites with fully periodic boundary conditions (PBC). The
occupation of each site is represented by a spin variable $s_{xy}$ taking
the values $+1,-1,0$, if a positive or negative charge or the vacancy is
present at site ($x,y$). For simplicity, we restrict our study to equal
densities of positive and negative charges. 
At each MC step (MCS), the vacancy exchanges with a
randomly chosen nearest neighbor $s_{xy}$, with Metropolis rate~%
\cite{metropolis}, $\min \left\{ 1,\exp (s_{xy}E\delta y)\right\} $. Here, $%
\delta y=0,\pm 1$ is the change of the particle's $y$--coordinate due to the
jump. $E$ denotes the bias, uniform in space and time and directed along
positive $y$.

%=============================Fig. 1================
\begin{figure}[tbp]
\input epsf \hfill \hfill 
\begin{minipage}{0.16\textwidth}
  \epsfxsize = \textwidth \epsfysize = 1.5\textwidth \hfill
  \epsfbox{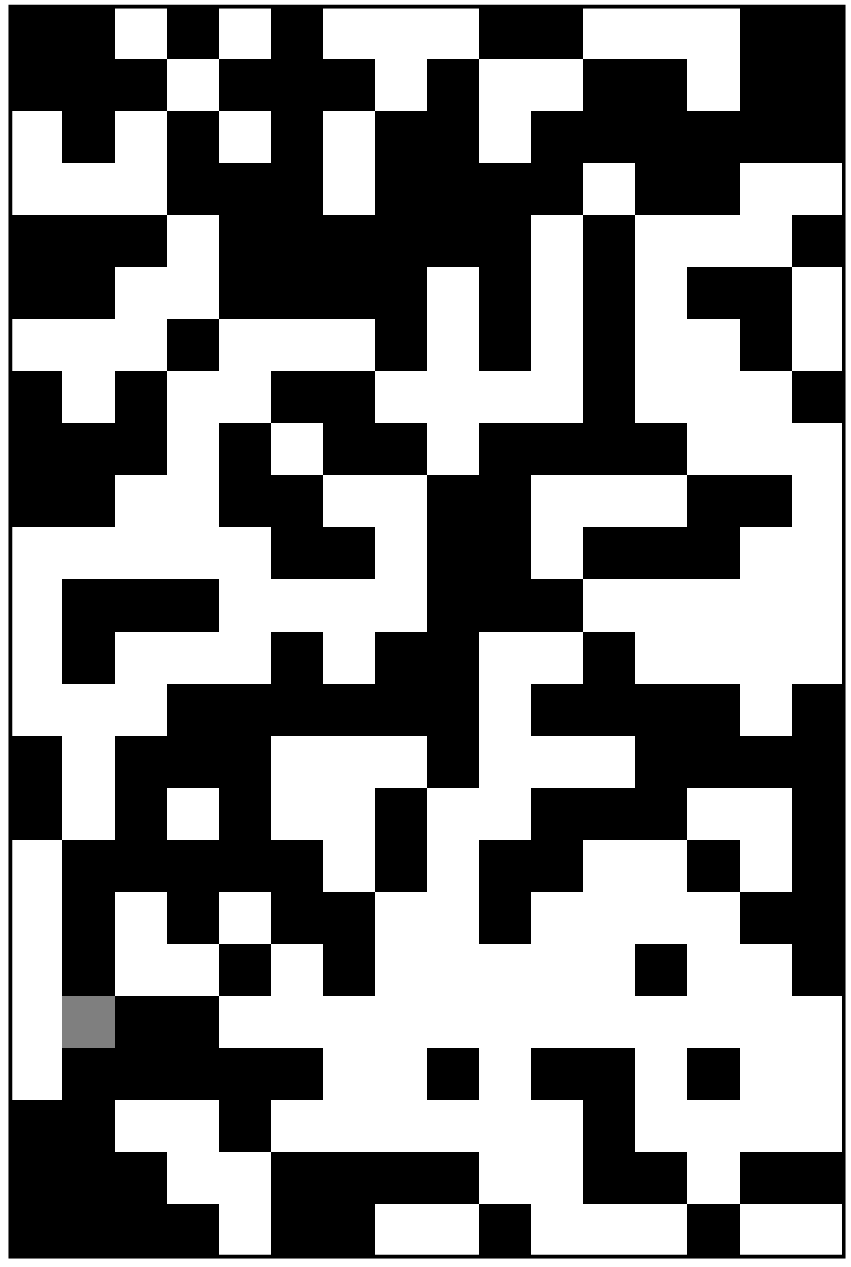} \hfill
    \begin{center} (a) \end{center}
  \end{minipage}
\hfill \hfill 
\begin{minipage}{0.16\textwidth}
  \epsfxsize = \textwidth \epsfysize = 1.5\textwidth \hfill
  \epsfbox{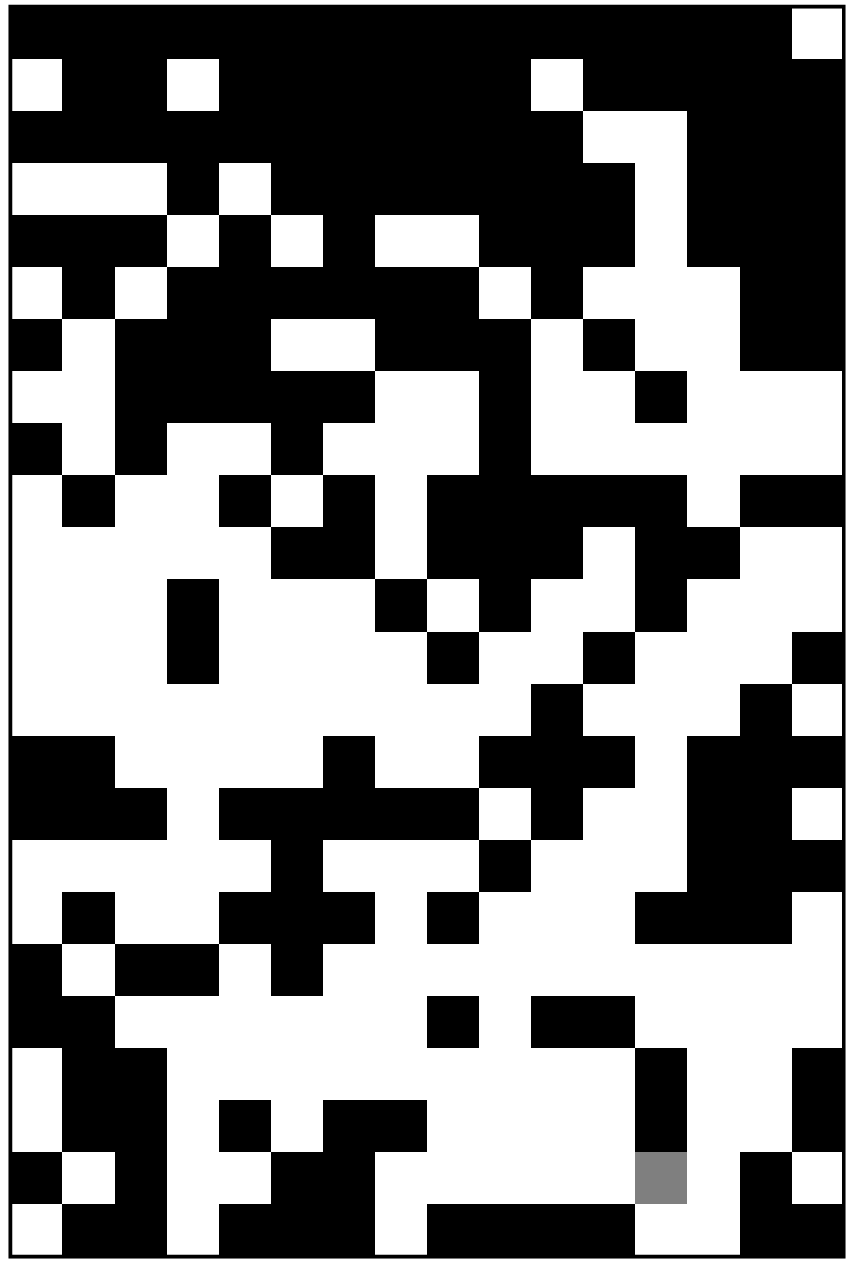} \hfill
    \begin{center} (b) \end{center}
  \end{minipage}
\hfill \hfill \vspace{0.02\textwidth} %\par
%\hspace{1.7cm}
\begin{minipage}{0.16\textwidth}
  \epsfxsize = \textwidth \epsfysize = 1.5\textwidth \hfill 
  \epsfbox{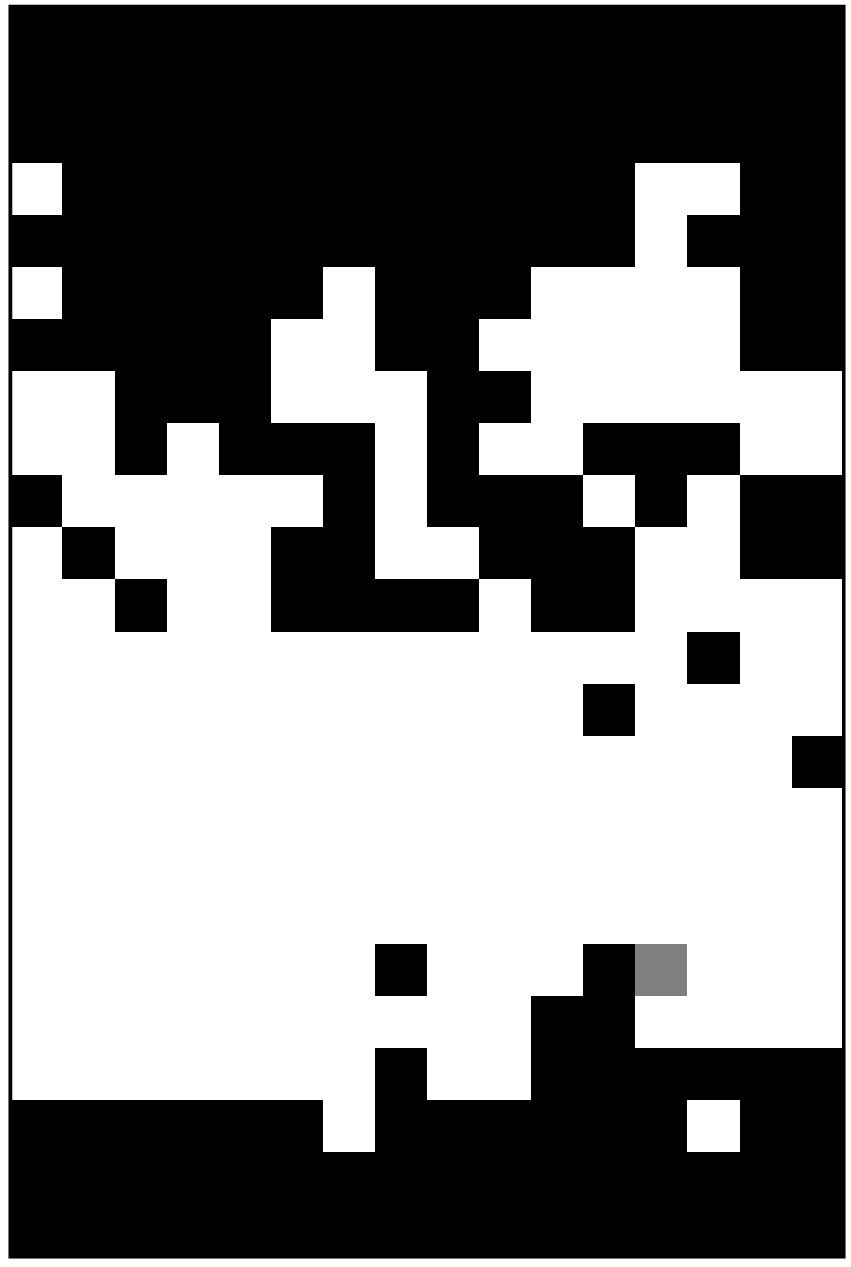}
    \begin{center} (c) \end{center}
  \end{minipage}
\hfill \hfill 
\begin{minipage}{0.16\textwidth}
  \epsfxsize = \textwidth \epsfysize = 1.5\textwidth \hfill
  \epsfbox{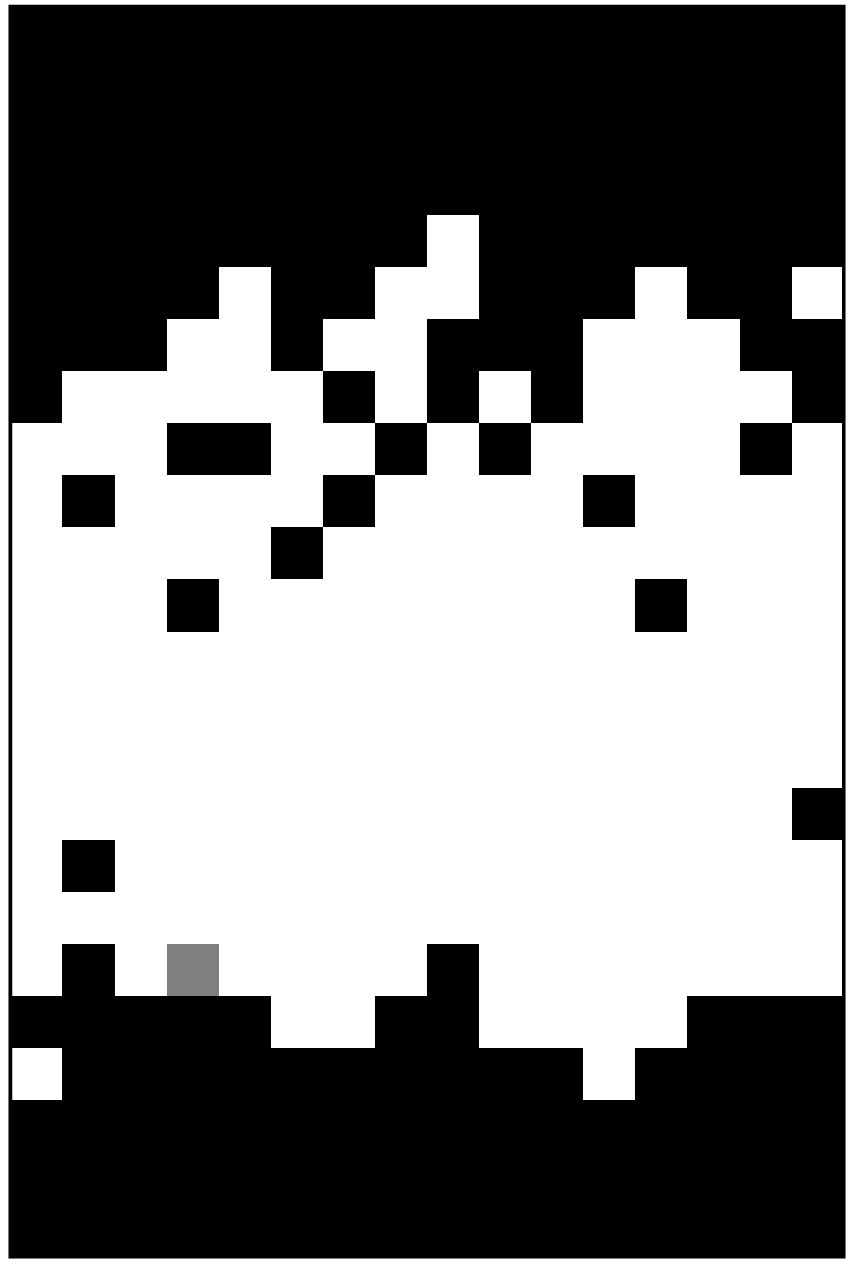}
    \begin{center} (d) \end{center}
  \end{minipage}
\hfill \hfill 
\begin{minipage}{0.16\textwidth}
  \epsfxsize = \textwidth \epsfysize = 1.5\textwidth \hfill
  \epsfbox{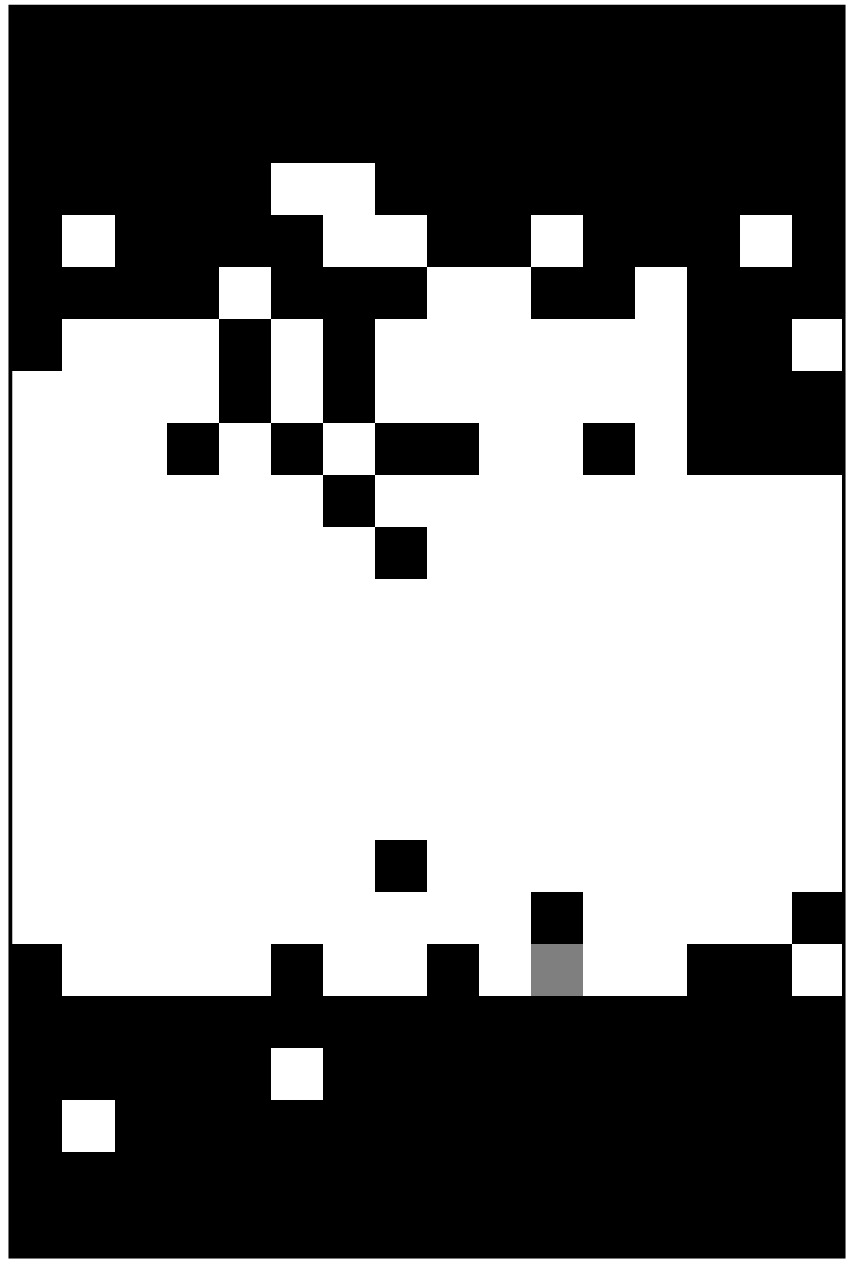}
    \begin{center} (e) \end{center}
  \end{minipage}
\hfill \hfill \vspace{-.4cm} 
\caption{Snapshots of a $16\times 24$ lattice at $E=0.8$, taken after 
(a) $10^{3}$, (b) $10^{4}$, (c) $10^{5}$, (d) $10^{6}$, 
(e) $10^{7}$ MC steps. 
Negative (positive) particles are colored black (white), and the
hole is gray. $E$ points upwards. The initial configuration was
random.}

\vspace{-.4cm}
\end{figure}
%=======================================================

This deceptively simple dynamics induces a \emph{%
charge segregation} process on the lattice, as illustrated by Fig.~1. 
Starting from a random
configuration (not shown), the system remains disordered for early times
(Fig.~1a). Eventually, the hole begins to segregate the two species: In 
Fig.~1b, an interface between regions of opposite charge begins to develop. Due
to PBC, a second interface must also form; this occurs at a time set by $%
L_{y}$. After $10^{5}$ MCS (Fig.~1c), the segregation of charges is already
quite apparent. Clearly, the field drives the hole \emph{towards }the lower
(the ``downstream''), and \emph{away from} the upper (the ``upstream''),
interface. Typical configurations are homogeneous in the transverse direction.
For our choice of parameters, the interfaces are well separated 
\cite{TS}. Since any increase of order requires a series of \emph{%
field-suppressed} exchanges, the approach to steady state is very slow
(Figs.~1c-e).

In the following, we provide a brief quantitative analysis of this process,
beginning with our MC results. 
The control parameters of our study are $E$, ranging from $0.2$ to $2.0$,
and the system size, $L_{x}=16$ or $40$, $16\leq L_{y}\leq 72$. 
The initial configuration is random. Time is measured in MCS. All of our data are averaged 
(denoted by $\left\langle \cdot \right\rangle $) over 
$100$ independent runs (samples), resulting in good statistics 
with errors well below $5\%$.

To probe the growth of the ordered domain, we measure averaged \emph{hole
and charge density profiles}, defined via $\phi (y,t){}\,\equiv
\,\left\langle \frac{1}{L_{x}}\sum_{x}(1-s_{xy}^{2})\right\rangle $ and $%
\psi (y,t)\,\equiv \,\left\langle \frac{1}{L_{x}}\sum_{x}s_{xy}\right\rangle 
$. Due to translational invariance, strips can be centered at any $y$; thus,
care must be taken when averaging: In each sample, we determine the maximum
of the hole density over a sufficient time interval. This maximum marks the
downstream interface. The charge density profiles from different samples are
now shifted so that these maxima coincide, and averages can be taken.
Clearly, this procedure is not very reliable for early times, since no or
only faint strips have developed yet; however, our focus here concerns later
stages of the growth process. Then, fluctuations of the downstream interface
position are rather small and very slow, so that its location can be
determined very accurately.

The evolution of the charge density profiles proceeds in three stages, as
illustrated in Fig.~2 for systems with different $L_{y}$. At early times,
the downstream interface forms and equilibrates quite rapidly, reaching its
steady state form at about $10^{5}$ MCS, independent of $L_{y}$. Near this
interface, the charges are already perfectly segregated, with the charge
profile saturating at $\pm 1$. Two fronts, one on each side, separate the
ordered from the remaining disordered region. The larger systems now enter a
second regime, during which both fronts travel slowly outwards, increasing
the width of the ordered region. This process is also independent of $L_{y}$%
: the centers of the fronts move out as $\ln t$, while their shape remains
unchanged. Eventually, in the third regime, the two fronts merge, due to
PBC, and the upstream interface equilibrates. The system has now reached
steady state. For smaller systems (including the one shown in Fig.~1), the
crossover to steady state sets in before well-developed fronts can form.

%====================Fig. 2====================

\begin{figure}[tbp]
\input{epsf} \vspace{-.5cm} \hspace{2.5cm} 
\begin{minipage}{.4\textwidth}
    \epsfxsize = 1.3\textwidth \epsfysize = \textwidth 
  \epsfbox{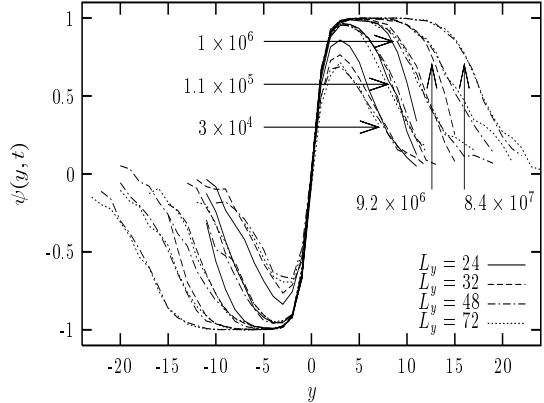} 
  \end{minipage}
\vspace{-.4cm} 
\caption{Charge density profiles for $L_{x}=16$, $E=0.8$, and a range of 
$L_{y}$, at different MC times (arrows). Some data
points, away from the central region, are not shown for clarity.}
\label{cd_Ly24324872_E08}
\vspace{-.5cm}
\end{figure}

%=============================================

As a quantitative probe,  
we define an order parameter, $\bar{Q}\equiv
\sum_{y}\psi ^{2}(y,t)$~\cite{TS} which measures the area, and hence the
degree of order, under the (averaged, squared) charge profile. A central
result of our work, Fig.~3 demonstrates that $\bar{Q}$ exhibits dynamic
scaling in the logarithmic growth regime: Excellent data collapse is
observed, for a range of $E$, $L_{x}$ and $L_{y}$ if $E\bar{Q}$ is plotted
vs the scaling variable $\tilde{t}\equiv tE^{3}/L_{x}$. This confirms our
picture of the ordering process: the motion and shape of the fronts is
independent of $L_{y}$, and $\bar{Q}(\tilde{t})\propto \ln \tilde{t}$, since
each escape of the vacancy is an activated process (exponentially suppressed
by $E$).

Two comments are in order. First, we should emphasize that $\bar{Q}$ differs
from an order parameter used previously, namely $Q\equiv
\left\langle \sum_{y}\left( \frac{1}{L_{x}}\sum_{x}s_{xy}\right)
^{2}\right\rangle $~\cite{SHZ}. The latter can be measured directly from the
configurations, without the shifting procedure. Essentially, $Q$
counts the \emph{total} number of ordered rows transverse to the external
field. It is therefore sensitive to the presence of \emph{multiple strips}
which can emerge during the ordering process, especially in larger systems.
As a result, the growth of $Q$ exhibits a weak $L_{y}$-dependence. In
contrast, the shifting procedure averages out multiple strips, in favor of
the largest (dominant) strip. Here, we focus on $\bar{Q}$ but note that the
study of multiple strips remains an outstanding problem. Second, our
scenario of the ordering process relies on having a well-developed
downstream interface sandwiched between two fully charge-segregated regions.
Our steady-state studies~\cite{TS} show that this is the case
provided $EL_{y}\geq 18$. Thus, the system with $L_{y}=24$, $E=0.8$ sets a
lower limit.

\emph{Analytic approach. }Finally, we turn to an analytic description, 
with the aim of deriving the appropriate scaling variables and the 
slope of the $\ln \tilde{t}$ law. 
Our starting point is a set of mean-field equations for the
(coarse-grained) local hole and charge densities. Due to particle number
conservation, the continuum version of our model takes the form of two
continuity equations which can be derived from a microscopic
master equation. Their form has been well established~\cite{SHZ,VZS,TS}.
Proceeding directly to the equations for the profiles, we obtain 
\begin{eqnarray}
\partial _{t}\phi (y,t) &=&\partial \left\{ \partial \phi +E\phi \psi
\right\}    \\
\partial _{t}\psi (y,t) &=&\partial \left\{ \phi \partial \psi -\psi
\partial \,\phi -E\phi \right\}   \label{prof}
\end{eqnarray}
Here, $\partial $ denotes a spatial derivative in the drive direction, and
terms of $O(\phi ^{2})$ have been neglected, since they reflect the presence
of multiple vacancies. The continuum limit has been taken by letting the
lattice constant vanish at finite $E$. The equations have to be supplemented
with appropriate boundary conditions and the constraints on total mass and
charge.

%===================Fig. 3===============================

\begin{figure}[tbp]
\input{epsf} \vspace{-.5cm} \hspace{2.5cm} 
\begin{minipage}{.4\textwidth}
    \epsfxsize = 1.3\textwidth \epsfysize = \textwidth 
  \epsfbox{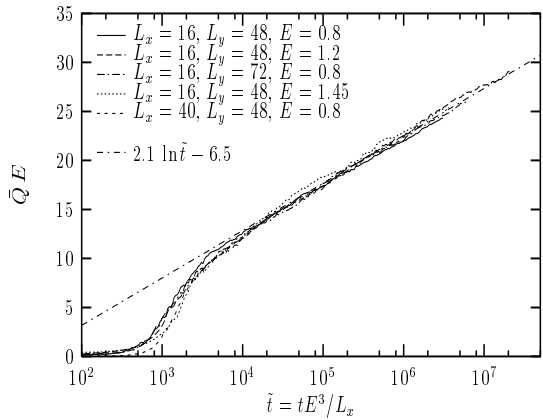} 
  \end{minipage}
\vspace{-.2cm}
\caption{Scaling plot of order parameter. The straight line, with slope
$2.1\pm0.1$, is
a fit to the data.}
\label{logfit}
\vspace{-.3cm} 
\end{figure}

%===========================================================

Solving these coupled, nonlinear PDE's is of course highly nontrivial.
However, the data suggest a few approximations which
allow for considerable analytic progress. Since we wish to capture 
the logarithmic growth regime, we
center the downstream interface at the origin ($y=0$) and focus on the front
moving in the positive $y$-direction (the other front being a mirror image,
of course).
By the onset of the second regime, this front is located 
well away from $y=0$, since 
the charge density at the downstream
interface has already equilibrated (cf. Fig.~2).
Moreover, MC data
show that the hole density profile is strongly localized at the origin and
does not change significantly as time progresses. Therefore, we replace it in Eq.~(\ref{prof})
by its \emph{steady state} form $\phi _{o}(y)$~\cite{TS} 
which decays, to excellent approximation, as 
$c^{-1}\exp (-E|y|)$ on both sides. Here, $c\propto L_{x}E^{-1}$ is a
normalization, such that $\phi _{o}(y)$ is the \emph{probability} 
of finding the vacancy at site ($x,y$) in the system. 
Introducing the scaling variables, 
$z\equiv Ey$ and $%
\tau \equiv c^{-1}E^{2}t$, we arrive~\cite{long} at a parameter-free equation,
valid for $y>0$: 
\begin{equation}
\partial _{\tau }\psi (z,\tau )=\exp (-z)\left[ \partial _{z}^{2}\psi -\psi
+1\right] .  \label{psi}
\end{equation}
Clearly, initial and boundary conditions are needed. To study 
the motion of the front well before saturation
occurs, we may allow $0<z<\infty $~\cite{c}. The disordered phase far ahead
of the front is modelled by $\lim_{z\rightarrow \infty }\psi (z,\tau )=0$
for all $\tau <\infty $. Further, the fully ordered phase well behind the
front should be independent of $t$, since it is unaffected by further
growth; thus, we demand that $\psi (z,\tau )\rightarrow 1$ for $\tau
\rightarrow \infty $ and all $0<z<\infty $.

Some comments are in order. First, since $c\propto L_{x}/E$, the
characteristic scaling of MC time with $E^{3}/L_{x}$ already emerges.
Second, since $\bar{Q}\equiv 2\int_{0}^{\infty }\psi
^{2}(y,t)dy=2E^{-1}\int_{0}^{\infty }\psi ^{2}(z,\tau )dz$, our mean field
theory predicts data collapse if $\bar{Q}E$ is plotted vs $tE^{3}/L_{x}$, as
borne out by Fig.~3.

Since the moving front retains its shape, we may seek a
solution in the form $\psi (z,\tau )\equiv f(w)$, with $%
w\equiv z-z_{o}(\tau )$. Clearly, $z_{o}(\tau )$ describes the front
position. Eq.~(\ref{psi}) becomes

\begin{equation}
-\dot{z}_{o}e^{z_{o}+w}f^{\prime }=f^{\prime \prime }-f+1,  \label{front}
\end{equation}
where $\dot{z}_{o}\equiv $ $\partial _{\tau }z_{o}$ and $f^{\prime }\equiv
\partial _{w}f$, with $\lim_{w\rightarrow +\infty }f(w)=0$ and $%
\lim_{w\rightarrow -\infty }f(w)=1$. Also, $f^{\prime }$ differs notably
from zero only in a (finite) neighborhood of $w=0$. Multiplying both
sides of (\ref{front}) by $f^{\prime }$, and integrating from $-\infty $
to $+\infty $, we arrive at $\dot{z}_{o}e^{z_{o}}\int_{-\infty }^{+\infty
}dwe^{w}f^{\prime 2}=\frac{1}{2}.$ Obviously, $\int_{-\infty }^{+\infty
}dwe^{w}f^{\prime 2}\equiv \kappa $ is a positive, nonvanishing \emph{%
numerical} constant. The time dependence of the front position (with simple
initial condition $z_{o}(0)=0$) now follows as $z_{o}(\tau )=\ln \left(
1+\tau /2\kappa \right) $. It is logarithmic, as expected.

Returning to Eq.~(\ref{front}), and using $\dot{z}_{o}e^{z_{o}}=1/2\kappa $,
the shape of the front, subject to the specified boundary conditions, is given by $f(w)=2\kappa e^{-w}\left[ 1-\exp (-e^{w}/2\kappa)\right]$. 
The associated order parameter is now easily computed: $\bar{Q}\,E=2\ln \tau
-0.927...+O(1/\tau )$, for large times $\tau >>\kappa $. Gratifyingly, the
key features of the MC data, namely, the correct scaling variables, the
logarithmic growth law and even its amplitude (i.e., the factor $2$) are
reproduced by our solution. Of course, a single additive constant is
required to match the time scale of the simulations.

\emph{Conclusions. }In this letter, we presented MC and analytic results for
domain growth in a simple driven lattice gas. A single vacancy rearranges
positive and negative particles into charge-segregated strips 
transverse to the drive. Focusing on the largest strip, 
we find that it grows logarithmically
with time, via field-suppressed excursions of the vacancy away from the
center of the strip. The ordered domain is separated from the remaining
disordered region by two well-defined moving fronts which retain their shape
during the growth process. Eventually, the two fronts merge due to our BC's,
and the system crosses over into steady state. During the logarithmic growth
regime, we observe dynamic scaling of a suitably defined order parameter, 
$\bar{Q}$, if $\bar{Q}E$ is plotted vs $tE^{3}/L_{x}$. These findings
are supported by 
an (approximate) solution of a set of mean-field equations for
the charge and hole density profiles. To leading order in $t$, 
the scaling function 
$\bar{Q}E$ grows as $2\ln \left( tE^{3}/L_{x}\right) $, 
in good agreement with the data.

Several questions remain open. First, multiple-strip configurations form
easily, especially in larger systems, and our focus on the dominant strip
washes out any secondary ones, due to the shifting procedure. 
Unfortunately, a full
dynamic scaling study of observables sensitive to multiple strips demands
much greater computational effort. Whether similar coarsening behavior
emerges remains to be explored. A second natural extension of our study
would allow for multiple vacancies, i.e., a finite hole density. Under these
conditions, particles first form ``clouds'' which coarsen and eventually
merge into multiple strips transverse to the field. These strips then
continue to coarsen until a single charge-segregated strip survives. A
numerical solution of the 1D mean-field equations~\cite{kertesz} for the
multiple-strip regime indicates that logarithmic growth persists, consistent 
with the arguments of Kafri, et.al.~\cite{2D-c}.
However, a detailed analysis, which identifies the relevant dynamic scaling
variables and functions, and incorporates the crossover from clouds to
strips, is still outstanding.

\acknowledgments

We thank E.~Sch\"{o}ll, R.K.P.~Zia, G.~Korniss, and Z.~Toroczkai for
valuable discussions. This research is supported in part by the National
Science Foundation through the Division of Materials Research.


\begin{thebibliography}{99}
\bibitem{c}  These boundary conditions are already implicit in the
normalization of $\phi _{o}(y)$.
\end{thebibliography}
\end{document}